\documentclass[aps,twocolumn,prresearch,english]{revtex4-1}

\usepackage{amsmath}
\usepackage{amssymb}
\usepackage{amsthm}

\usepackage{amsbsy}
\usepackage{amstext}
\usepackage{graphicx}
\usepackage[dvipsnames]{xcolor}
\usepackage{float}

\usepackage{wasysym}

\usepackage{mathtools}

\makeatletter
\usepackage{amsfonts}
\usepackage{dcolumn}
\usepackage{bbold,bm}

\newcommand\vex[1]{\mathbf{#1}}
\newcommand\gvex[1]{\boldsymbol{#1}}

\def\id{\mathbb{1}} 

\def\highlight#1{#1}

\makeatother

\begin{document}

\title{Floquet-Engineered Topological Flat Bands in Irradiated Twisted Bilayer Graphene}

\author{Yantao Li}
\affiliation{Department of Physics, Indiana University, Bloomington, Indiana 47405, USA}

\author{H. A. Fertig}
\affiliation{Department of Physics, Indiana University, Bloomington, Indiana 47405, USA}

\author{Babak Seradjeh}
\affiliation{Department of Physics, Indiana University, Bloomington, Indiana 47405, USA}

\begin{abstract}
We propose a tunable optical setup to engineer topologically nontrivial flat bands in twisted bilayer graphene under circularly polarized light. Using both analytical and numerical calculations, we demonstrate that nearly flat bands can be engineered at small twist angles near the magic angles of the static system. The flatness and the gaps between these bands can be tuned optically by varying laser frequency and amplitude. We  study the effects of interlayer hopping variations on Floquet flat bands and find that {changes associated with} lattice relaxation favor their formation. Furthermore, we find that, once formed, the flat bands carry nonzero Chern numbers. We show that at currently known values of parameters, such topological flat bands can be realized using circularly polarized UV laser light. Thus, our work opens the way to creating optically tunable, strongly correlated topological phases of electrons in moir\'e superlattices.
\end{abstract}


\maketitle


\section{Introduction}
Despite the simplicity of its structure, graphene and its multilayers have proven to support
a remarkable diversity of electronic behaviors~\cite{Castro_Neto_2009,Abergel_2010,Kotov_2012}.  Among such
systems, twisted bilayer graphene (TBG) has shown some of the most surprising properties.  When
rotated relative to one another, two graphene layers form a moir\'e pattern, which, even in the
absence of true commensuration, may be treated to a good approximation as a Bravais lattice
with a large unit cell~\cite{dosSantos_2007,Shallcross_2008,Shallcross_2010,Mele_2010,Mele_2011,Bistritzer_2011,dosSantos_2012}.
The possibility of unusual electronic states in this system has long been appreciated, in part
because the electronic structure supports van Hove singularities at relatively low energy~\cite{Li_2009,McChesney_2010,DeGail_2011,Nandkishore_2012,Luican_2012,Yan_2012,Brihuega_2012,Lu_2014a,Lu_2014b,
Kim_2016,Chung_2018}.
More recently, the potential for these systems to support extraordinarily flat bands at ``magic''
twist angles~\cite{Bistritzer_2011,dosSantos_2012,SanJose_2012} has been verified experimentally, and the
system demonstrated to display interaction physics in the forms of Mott insulating behavior and superconductivity~\cite{Cao_2018a,Cao_2018b}. The implications of this single-particle structure for collective electron
states is now an area of intense investigation~\cite{Po_2018,Guo_2018,Dodaro_2018,Huang_2018,Xu_2018,Liu_2018,
Ochi_2018,Wu_2018,Padhi_2018,Wu_2019,Zhang_2019a,Lian_2020,Lu_2019,Koshino_2019,Hejazi_2019b,Gonzalez_2019,Kang_2019,Fidryskiak_2018,Zhang_2019b}. 
While there has been significant progress, a complete
understanding of
the physical origin of the magic angle flatness remains elusive~\cite{Suarez_Morell_2010,Nam_2017,Zou_2018,Yuan_2018,Lin_2018,Zhang_2019,Pal_2018,Kang_2018,Koshino_2018,Rademaker_2018,Tarnopolsky_2019,Hejazi_2019a,Po_2019,Qiao_2018,Carr_2019,Guinea_2019}
In particular, magic angles are obtained at mechanically fine-tuned values, which cannot be changed once a sample is prepared~\cite{Yankowitz_2019}.

\begin{figure}[t]
   \centering
   \includegraphics[width=3.4in]{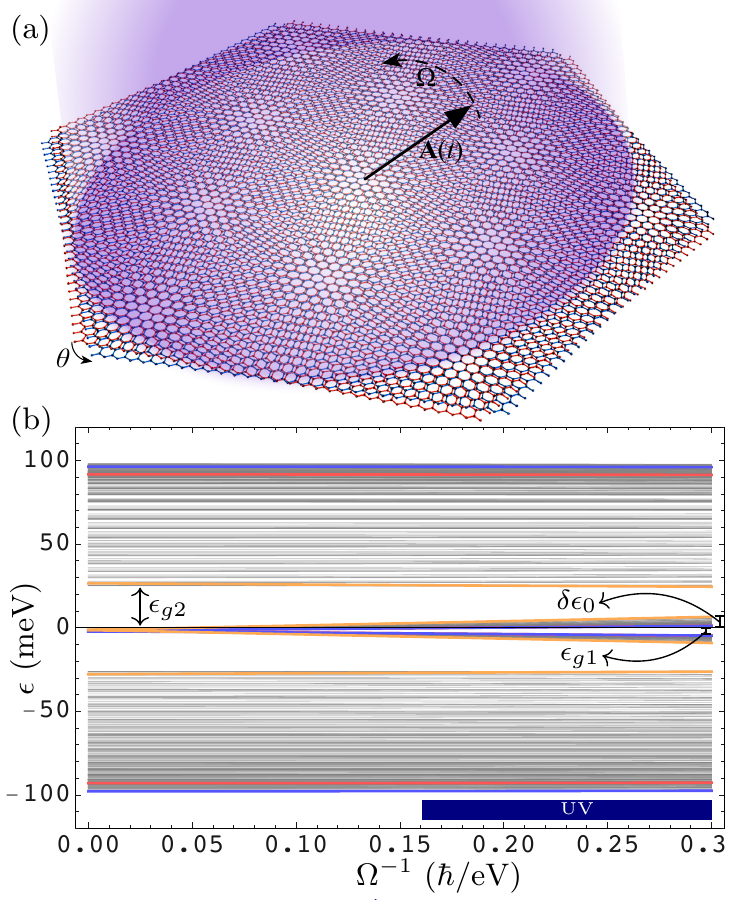} 
   \caption{(a) The setup of the twisted bilayer graphene irradiated by circularly polarized laser. (b) The four central bands vs. laser frequency $\Omega$ (the UV range $3.3~\text{eV}<\hbar\Omega<6~\text{eV}$ is marked by the horizontal bar) for twist angle $\theta\approx1.1^{\circ}$, 
   interlayer tunneling amplitudes $w_{AB}=w_{AA}/0.816 = 112$~meV, and laser field $E/\Omega = 65.8~\text{V}/c$. 
   The central flat bands are gapped and carry Chern numbers $\pm 4$.}
   \label{fig:setup}
\end{figure}

In this work, we demonstrate that TBG as a platform for correlated electronic
states is enriched by circularly polarized light impinging on the system. In particular, this offers a highly controllable optical setup to engineer topological flat bands over a {range} of small twist angles away from the magic angle, and allows \emph{in situ} optical tuning of the flat band structure in TBG. Moreover, we show that these Floquet flat bands are separated from each other and other bands by sizable energy gaps, and have nonzero Chern numbers. This combination of properties---extreme flatness, gaps, and non-vanishing Chern numbers---makes these bands near-perfect analogs of Landau levels, the energy spectrum of two-dimensional electrons in
a static magnetic field.  This offers the potential that electrons in this setting
can support states akin to those
found in the fractional quantized
Hall effect~\cite{Yoshioka_book,Jain_book}, stabilized by repulsive interactions among the electrons.  Indeed, such ``Floquet fractional Chern insulators'' have been argued to be present
in honeycomb lattices subject to circularly polarized light~\cite{Grushin_2014}.
As we explain below, TBG is special in that the large size of the moir\'e
unit cell allows the flat Chern band to emerge at relatively low excitation
energies relative to that needed to induce topological behavior in more microscopic
honeycomb lattices such as single layer graphene~\cite{Kundu_2014,Kundu_2016}. (This possibility was recently explored at larger twist angles~\cite{Topp_2019}, where no flat bands were observed.) Thus, our work shows that irradiated TBG represents an extremely attractive setting
to search for fractional Chern insulators and other correlated electron states.

\section{Setup and model}
Our setup is illustrated in Fig.~\ref{fig:setup}(a). A circularly polarized laser beam with vector potential $\vex A(t) = A (\cos \Omega t, \sin \Omega t)$ and frequency $\Omega$ is directed normal to the TBG.
In the context of single layer graphene and related systems, time-dependent perturbations have been found to modify the
electronic properties in ways that allow behaviors inaccessible to
the bare materials to emerge~\cite{Oka_2009,Abergel_2009,Topp_2019,Gu_2011,Kitagawa_2011,Morrell_2012,Dora_2012,Iadecola_2013,
Cayssol_2013,Ezawa_2013,Katan_2013,Syzranov_2013,
Reynoso_2013,Lindner_2013,Rechtsman_2013,Wang_2013,Kundu_2014,Fregoso_2014,Farrell_2015,Dehgani_2015,Kundu_2016,Farrell_2016,Roy_2017,Yao_2017,Kolodrubetz_2018,Crowley_2020,Rudner_2020,McIver_2019}. Provided the driving potential does not
break the translational symmetry of the lattice, the spectrum can be labeled by the crystal momentum ${\bf k}$, which falls within the same Brillouin zone as that of the static system. The dynamics of the temporally periodic Hamiltonian $H_{\vex k}(t) = H_{\vex k}(t+2\pi/\Omega)$ is governed by the Floquet-Schr\"odinger equation $[H_{\vex k}(t) - i\hbar\partial_t]|\phi_{\vex k s}(t)\rangle = \epsilon_{\vex k s} |\phi_{\vex k s}(t)\rangle$,
and supports quantum states characterized by the Floquet bands of quasienergies $\epsilon_{\vex k s}$, labeled by $s$, whose values fall into a ``Floquet zone'' of size $\hbar\Omega$. The corresponding Floquet modes $|\phi_{\vex k s}(t)\rangle$ are also periodic and furnish a basis for the solutions of the time-dependent Schr\"odinger equation, $|\psi_{\vex k s}(t)\rangle \equiv e^{-i \epsilon_{\vex k s} t /\hbar} |\phi_{\vex k s}(t)\rangle$.

For sufficiently large frequency, we can model the low-energy dynamics of irradiated twisted bilayer graphene following Bistritzer and MacDonald~\cite{Bistritzer_2011}
and treat the interlayer hopping in an effective long-wavelength approximation (\highlight{see the Appendix for a discussion of the validity of the model in the presence of periodic drive}). In the position basis,
\begin{equation}\label{eq:FTBG}
H(t) = \begin{bmatrix}
h_{\theta/2}(t) & T \\
T^\dagger & h_{-\theta/2}(t)
\end{bmatrix},
\end{equation}
where $h_{\theta/2}(t) = v_F[-i\hbar \gvex\nabla - e \vex A(t)]\cdot\gvex\sigma_{\theta/2}$, with rotated Pauli matrices $\gvex\sigma_{\theta/2} \equiv e^{-i\theta\sigma_z/4} (\sigma_x,\sigma_y) e^{i\theta\sigma_z/4}$ and Fermi velocity $v_F$, is the low-energy Dirac Hamiltonian of one of the valleys of a single graphene sheet twisted by angle $\theta/2$. The interlayer tunneling matrix $T = \sum_{n=1}^3 T_n e^{-i k_\theta \vex q_n \cdot \vex r}$, with
\begin{equation}
T_n = w_{AA} \sigma_0 + w_{AB} \vex q_n \cdot \gvex\sigma_{\pi/2},
\end{equation}
where the unit vectors $\vex q_1 = (0,-1)$, $\vex q_{2,3} =  (\pm\sqrt3/2,1/2)$ encode the tunneling $w_{AA}$ and $w_{AB}$ between the AA- and AB-stacked regions of the twisted bilayer graphene. Here, $k_\theta = 8\pi \sin(\theta/2)/3 a$ sets the wavevector of the moir\'e pattern and $a$ is the Bravais lattice spacing of graphene.

In our numerical calculations, we have taken parameter values $a= 2.4$~\AA, $\hbar v_F / a = 2.425$~eV, and $w_{AB} = 112$~meV based on experimental observations. We study the Floquet spectrum and compare to the static situation as a function of $u \equiv w_{AA}/w_{AB}$, the twist angle $\theta$ [or equivalently $\alpha \equiv w_{AB}/{\hbar v_F k_\theta} = 1.1\times 10^{-2} / 2\sin(\theta/2)$], the laser frequency $\Omega$ and the electric field amplitude $E = \Omega A$.

\section{Topological Floquet flat bands}
For sufficiently high frequencies, we may find the Floquet spectrum from an effective static Floquet Hamiltonian $H_F \approx \overline H + \delta H_F$, where $\overline H = H^{(0)}$, $\delta H = [H^{(-1)},H^{(1)}]/\hbar\Omega$, and $H^{(n)} = \int_0^1 e^{-2\pi in \tau}H(2\pi\tau/\Omega) d\tau$ are the Fourier components of the periodic Hamiltonian. In our setup, $\overline H$ is the Hamiltonian of the static system found by setting $\vex A=0$, which supports flat bands at a series of magic angles. 
The leading term at high frequencies is spatially uniform and is given by
\begin{equation}
\delta H_F = \frac{(ev_F A)^2}{\hbar \Omega} \sigma_z\otimes \id.
\label{sigzterm}
\end{equation}
{Note that $v_F$ here is the Fermi velocity of single layer graphene, which is
considerably larger than the Fermi velocity for Dirac points of the static moir\'e lattice under flat band conditions. This relatively large coefficient allows non-trivial physics in the
irradiated system to emerge with only moderate laser amplitudes.}

For $u = 0$, $\overline H$ is chirally symmetric under the chiral operator $\sigma_z\otimes \id$; thus, $\{\overline H,\delta H_F\} = 0$. Since the zero energy states of $\overline H$ can be chosen to be eigenstates of the chiral operator, they remain eigenstates of $H_F$ in the presence of $\delta H_F$ but acquire a finite energy $\pm (ev_FA)^2/\hbar\Omega$. In the chiral limit, $\overline H$ has two degenerate absolutely flat bands~\cite{Tarnopolsky_2019}, $\overline H |\psi_{\vex k \pm}\rangle = 0$. Therefore, the Floquet spectrum will also have two absolutely flat bands with a finite central gap $\epsilon_{g1} = 2 (e v_F A)^2/\hbar \Omega$. For $u>0$, this gap is modified but we expect that it will be nearly linear in  $\Omega^{-1}$ for large frequencies. This is indeed what we find in our numerical solutions shown in Fig.~\ref{fig:setup}(b).

In contrast to the static situation, these Floquet flat bands have nonzero Chern numbers. One way to see this is that $\delta H_F$ acts \emph{the same} way on both layers, so that the gap at the moir\'e $K$ and $K'$ points have the same pattern of time-reversal symmetry breaking. Therefore, the total Chern number of the gapped flat bands must be nonzero. A more explicit way to show this result is to use the solutions for the absolute flat bands in the chiral limit $u=0$~\cite{Tarnopolsky_2019}. The $\vex k$ dependence of the entire band is given by the same Siegel theta function, confirming the same gap is produced at moir\'e $K$ and $K'$ points. Since these points are Dirac points of the central bands for all values of $\alpha$ (with zero velocity at the magic angles), the total Chern number is the sum of the two Chern numbers of the gapped Dirac points, i.e. $C_\pm = \pm(\frac12+\frac12) = \pm1$.

Due to time-reversal symmetry breaking by the circularly polarized laser field, the Chern number at the other valley of the single layer graphene must also have the same sign. To see this concretely, one may note that an appropriate Bistritzer-MacDonald model for the other valley may be obtained by an inversion ${x} \mapsto -{x}$ in Eq.~(\ref{eq:FTBG}), or, equivalently, by changing the sign of terms proportional to $\sigma_x$ prior to the $\pm \theta/2$ rotation.  This has the effect of changing the sign of $\delta H_F$  generated at high frequency in Eq.~(\ref{sigzterm}), thus producing a Floquet gap at the other valley with the opposite sign. This sign reversal is a consequence of time-reversal breaking. For a momentum path surrounding a moir\'e $K$ or $K'$ point, the line integral of the Berry's connection also receives an extra minus sign due to spatial inversion. The net effect of this is to induce the same Chern number in the moir\'e band of the graphene at the other valley. Thus, including the spin degeneracy factor, the central Floquet bands acquire a nontrivial Chern number $C=\pm 4$.

The Chern numbers are stable away from the chiral limit $u>0$ as long as the bulk gap remains open. As we show below, there is indeed a finite gap over a wide range of parameters which, in particular, includes the range of experimentally relevant values.


\section{Numerical results}
\begin{figure}[t]
  \centering
   \includegraphics[width=3.45in]{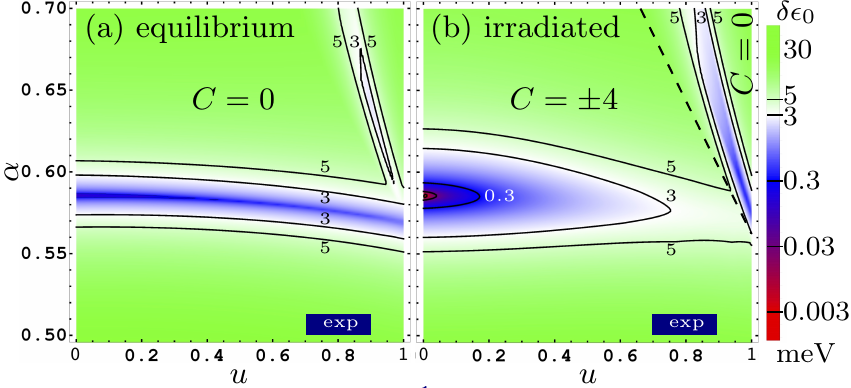} 
   \caption{The width of the central positive band of (a) static and (b) irradiated twisted bilayer graphene as a function of twist angle and the ratio of interlayer tunneling $u=w_{AA}/w_{AB}$. The range of $u$ in current experiments is shown by the horizontal bar. The central bands of the irradiated system around the magic angle are nontrivial and have Chern numbers $C=\pm4$, but become trivial across a gap closing shown by the dashed line. The laser frequency and electric field are set at $\hbar\Omega=6$~eV and $E = 2\times10^{4}$~kV$/$cm.
   }
   \label{fig:bandwidth}
\end{figure}

We have solved the Floquet-Schr\"odinger equation for our model Hamiltonian~(\ref{eq:FTBG}) and calculated the Chern numbers~\cite{Fukui_2005} associated with the Floquet bands numerically. In Fig.~\ref{fig:bandwidth}(a), we show the bandwidth of the two central bands for the equilibrium (static) and the irradiated system for laser frequency $\hbar\Omega = 6$~eV and electric field $E = 2\times10^4$~kV$/$cm. For the whole range of interlayer tunneling ratio $0\leq u\leq 1$, the static system shows flat bands at magic angle close to the chiral limit ($u=0$) value $\theta \approx 1.08^\circ$ corresponding to $\alpha\approx0.586$. We also note the appearance of flat bands at lower twist angles $\theta\approx 0.93^\circ$ ($\alpha\approx 0.68$) in a range $u>0.8$ that are unrelated to those appearing in the chiral limit.
Interestingly, evidence of such flat bands has recently been observed in experiment~\cite{Codecido_2019}.

For the irradiated case illustrated in
Fig. \ref{fig:bandwidth}(b),
two central Floquet flat bands, formed around the value of the magic angle in the chiral limit,
are apparent. Several trends are noteworthy in this case: (i) Flat bands are observed over a wider range of twist angles, which, in the chiral limit, we estimate to be $1.01^\circ \lesssim \theta \lesssim 1.13^\circ$. (ii) These bands become even flatter for smaller values of $u$, leading to ultraflat bands, with bandwidths smaller than those in the equilibrium case by one or two orders of magnitude, over a range $u\lesssim0.2$. (iii) The flat bands at the lower twist angle (larger $\alpha$) are also narrower than their static counterparts. (iv) The bands for smaller $\alpha$ have stable Chern numbers
$C = \pm1$ per valley and per spin (total Chern number $C=\pm4$) over a wide region, including the chiral limit; however, the flat bands at larger $\alpha$ are trivial
due to a gap closing between the two regions.

\begin{figure}[t]
  \centering
   \includegraphics[width=3.45in]{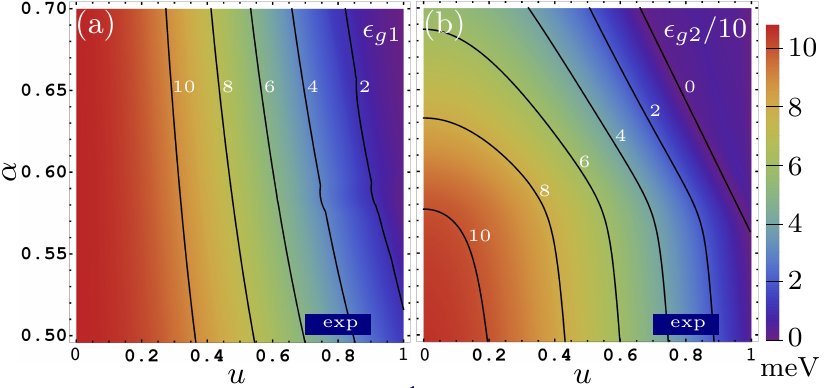} 
   \caption{Floquet band gaps between (a) central bands, $\epsilon_{g1}$, and (b) central positive band and the next band, $\epsilon_{g2}$, as a function of twist angle and the ratio of interlayer tunneling $u$. The laser parameters are the same as in Fig.~\ref{fig:bandwidth}.}
   \label{fig:bandgap}
\end{figure}

Band gaps for the Floquet spectrum are illustrated in Fig.~\ref{fig:bandgap}. The gap between the two central Floquet bands, $\epsilon_{g1}$, remains non-vanishing in the entire range of parameters shown. The gap $\epsilon_{g2}$ between the central bands and the next Floquet band is about an order of magnitude larger than $\epsilon_{g1}$, which means the central Floquet bands can be taken as reasonably flat and isolated. Note that the $\epsilon_{g2}$ gap closes for smaller twist angles just before the second flat band region appears.

\section{Experimental considerations}
In contrast to the static case, irradiated TBG offers an electron platform in which a filled flat band has non-trivial topology, in principle allowing for analogs of fractional quantum Hall states---fractional Chern insulators~\cite{Grushin_2014}---supporting quasiparticles of fractional charge and statistics.  The parameters needed to produce this platform should be within experimentally
realizable parameters.  For example, both the frequencies in the UV range and dynamical electric fields $\sim 10^4$~kV$/$cm used in our calculations are in ranges that are accessible by currently available laser technology~\cite{Fahad}.
{We also note that the work function of pristine graphene $\sim $ 4.5-5.1~eV~\cite{Song_2012,Klein_2018}, is above the low end of the range of frequencies we consider, so that our analysis in which photoemission is not included is appropriate.}

Our results are found over a range of
the parameter $u$, which physically represents the ratio of tunneling amplitudes between layers for sites on the same sublattice (AA tunneling) to ones on opposite sublattices (AB tunneling).  Because the lowest energy configuration for graphene bilayers involves AB stacking, a lattice-relaxed system will naturally have larger regions of AB alignment relative to AA alignment, suggesting $u<1${~\cite{Nam_2017,Koshino_2018,Tarnopolsky_2019,Carr_2019}}.  Fits to experimental data~\cite{Cao_2018a,Cao_2018b} yield $0.7<u<0.9$ (shown by horizontal bars in Figs.~\ref{fig:bandwidth} and~\ref{fig:bandgap}). In this range,
the width of central the Floquet bands near the largest magic angle ($\theta \approx 1.08^{\circ}$) is $\delta\epsilon_0 \approx 3$~meV,
while the second region of narrow bands at $\theta \lesssim 1^\circ$ has a width $\delta\epsilon_0 \approx 2$~meV.  Note that both these widths are smaller than our calculated width for the static system near $\theta \approx 0.93^\circ$, for which clear signatures of interacting flat band physics have been observed~\cite{Codecido_2019}.

The possibility of observing interacting topological flat band physics at the largest magic angle is further supported by the relative isolation of flat bands from bands
above and below; for $u\sim 0.8$, the gap between the central bands is of order
$\epsilon_{g1} \approx 4$meV, while the separation from bands above and below are of
order $\epsilon_{g2}\approx 30$~meV.  Thus experiments at temperatures at or below $\sim$~10 K
will avoid thermal excitations into these bands. Near the magic angle, assuming a dielectric constant $\kappa \sim 4$, we estimate the Coulomb energy for one electron per moir\'e unit cell as $3e^2 k_\theta/(4\pi)^2 \kappa \epsilon_0 \sim 44$~meV, hence the dominant energy scale in topological Floquet flat bands.

A striking feature of our results is that the situation improves notably as $u$ decreases, as is apparent in Figs. \ref{fig:bandwidth} and \ref{fig:bandgap}.
It should be possible to adjust $u$ by judicious choices of substrate or by application of pressure~\cite{Yankowitz_2019}.
Both can increase the interlayer
coupling, therefore increasing the sizes of the lower energy AB regions at the expense of the AA regions.
At small $u$, the bandwidth of the central Floquet bands can become extremely small, falling
well below that of the static system at the same parameters.  Moreover, the range of angles
for which the bandwidth is anomalously small increases notably, which in principle relaxes some of
the challenges associated with producing samples with finely tuned twist angles.

Static TBG has already proven to be a remarkable host for correlated, interacting electron physics.  The application of time-dependent fields offers a way of further enriching this system,
by introducing
non-trivial topology into the flat band.  Thus, irradiated TBG may well prove to be a unique
host for exotic, gapped quantum electron states with unusual quasiparticles, without the need for
magnetic fields to stabilize them.

\emph{Note added.} After the submission of this work, similar results were reported independently in Ref.~\cite{Katz_2020}.

\begin{acknowledgments}
The authors acknowledge useful conversation with Fahad Mahmood. This work is supported in part by the NSF through the CAREER
award DMR-1350663, as well as via grant Nos.
DMR-1506263, DMR-1914451, and ECCS-1936406. Further support was supplied
by the US-Israel Binational
Science Foundation grant No. 2016130, and the College of Arts and Sciences at Indiana University.
HAF acknowledges the support of the Research Corporation for Science Advancement through a Cottrell SEED Award.
We acknowledge the hospitality and support of the Max-Planck Institute for the Physics of Complex Systems in Dresden, and the Aspen Center for Physics via NSF grant PHY-1607611, where parts of this work were performed.
\end{acknowledgments}

\appendix*

\section{Validity of the Low-Energy Theory}
In this work, a linear approximation for the individual graphene layers is assumed, as in the original BM model. This approximation could, in principle, be relaxed. However, since we are interested in the low-energy bands of the twisted system in the range of energies below 100 meV, and since the laser frequencies we consider are $> 30$ times larger, the linear approximation is appropriate for our purposes.

Another consideration in a drive system is possible Floquet transitions at the Floquet zone edge, which may affect, among other things, the Chern numbers of the bands~\cite{Kundu_2014}. However, in twisted bilayer graphene the bands are already highly folded due to the small size of the moir\'e Brillouin zone, and the resulting band structure and band gaps are very different from those in single-layer graphene. The frequency we consider is a factor of $\sim 100-1000$ larger than the relevant energy scales of the low-lying bands of the twisted bilayer graphene. Since the central Floquet bands are gapped (both with themselves and with the higher bands) over a wide range of frequencies, any changes of Chern numbers associated with the Floquet zone edge transitions are not relevant for the low-energy central bands.

\begin{figure*}[t]
  \centering
   \includegraphics[width=7in]{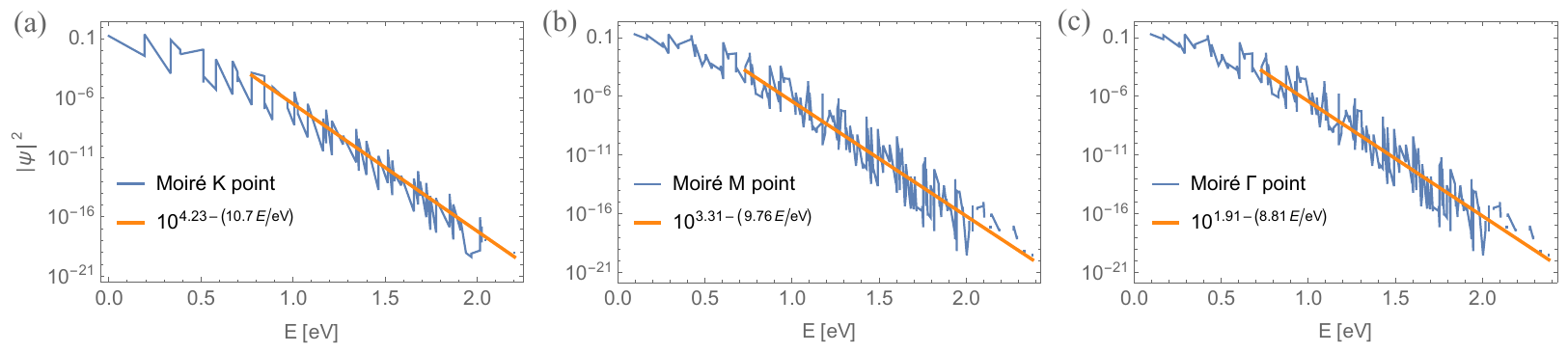} 
   \caption{The weight of the wavefunction of the central flat band of static twisted bilayer graphene vs. the energy of original graphene bands at the moir\'e (a) K , (b) $M$, and (c) $\Gamma$ points for $\alpha=0.58$ (at magic angle) and $u=0.816$ in the experimental range. The lines show exponential fits.}
   \label{fig:FloquetSB}
\end{figure*}

Finally, we consider the effect of Floquet sidebands on the low-energy physics. Since the Floquet sidebands occur at $\hbar\Omega \gtrsim 3$~eV, far removed from the low-energy central bands with energies $< 100$~meV, and in view of the large number of foldings into the moir\'e Brillouin zone at small twist angles, we expect that any admixture with the Floquet sidebands should be tiny. To estimate this effect, we calculate the admixture of high-energy states of pristine graphene into the flat bands of a static twisted graphene bilayer near its magic angle. In Fig.~\ref{fig:FloquetSB}, we show the weight of the wavefunction of the central flat band as a function of energy, at the moir\'e $K$, $M$, and $\Gamma$ points. Other points in the moir\'e Brillouin zone behave quite similarly, as do results for other values of $u$. The important features to note are that the weights are well-converged with respect to our cutoff, and are extremely tiny at high energies. Indeed, beyond $\sim 2.4$~eV, the wavefunction vanishes within our machine accuracy. However, fitting the available nonzero data points and extrapolating to our lowest drive frequency of $3$~eV, we find a weight of order $10^{-28}$, $10^{-26}$, and $10^{-25}$ respectively at moir\'e $K$, $M$, and $\Gamma$ points.

These weights are so small because they drop exponentially with the number of interlayer tunneling matrix elements to fold the high energy states of pristine graphene into the moir\'e Brillouin zone. Since the latter is $\sim 10^4$ times smaller than the Brillouin zone of a single graphene layer at the magic angle, the number of foldings needed is of the order of 100. While it is true that adding the temporal drive field will formally fold a static band at energy $\Omega$ down to zero quasienergy, its mixing with the flat bands is determined by the weight of the wavefunction and remains exceedingly small. This is so because the laser field is uniform in space. Therefore, this mixing requires the same number of interlayer tunneling matrix elements as in the static case to bring the wavevector content of the flat band in coincidence with that of the high-energy band.  A simple estimate of the resulting gap from a crossing between the flat band and the folded high-energy band is then $\sim |\psi|^2 (ev_FA)^2/(\hbar\Omega) \sim 3\times10^{10} |\psi|^2$~Hz~$< 10^{-26}$~meV, where we assumed $u\approx0.8$, electric field $\approx 4 \times 10^4$~V/cm, and $\hbar\Omega =3$~eV. Thus, Floquet sidebands and their admixture with the topological flat band are practically undetectable in this range of frequencies.


%

\end{document}